\documentclass[prl,twocolumn,superscriptaddress]{revtex4-1}
\usepackage[caption=false]{subfig}
\usepackage{amsfonts}
\usepackage{amssymb}
\usepackage{amsmath}
\usepackage{commath}
\usepackage{graphicx,bm}
\usepackage{verbatim}

\makeatletter
\newcommand*{\balancecolsandclearpage}{%
  \close@column@grid
  \clearpage
}
\makeatother

\newtheorem{theorem}{Theorem}

\newenvironment{proof}[1][Proof]{\noindent \textbf{#1: }}{\ \rule{0.5em}{0.5em}}

\def\avg#1{\langle#1\rangle}

\def\be{\begin{equation}} \def\ee{\end{equation}}
\def\bea{\begin{eqnarray}} \def\eea{\end{eqnarray}}

\begin{document}
\title{Exact Results on Itinerant Ferromagnetism and the 15-puzzle Problem}
\author{Eric Bobrow}
\address{Department of Physics and Astronomy, Johns Hopkins University,
Baltimore, Maryland 21218, USA}
\author{Keaton Stubis}
\address{Department of Mathematics, Johns Hopkins University, Baltimore,
Maryland 21218, USA}
\author{Yi Li}
\address{Department of Physics and Astronomy, Johns Hopkins University,
Baltimore, Maryland 21218, USA}
\date{April 5, 2018}

\begin{abstract}
We apply a result from graph theory to prove exact results about itinerant ferromagnetism.
Nagaoka's theorem of ferromagnetism is extended to all non-separable graphs except single polygons with more than four vertices by applying the solution to the generalized 15-puzzle problem, which studies whether the hole's motion can connect all
 possible tile configurations.
This proves that the ground state of a $U\to\infty$ Hubbard model with one hole away from the half filling on a 2D honeycomb lattice
or a 3D diamond lattice is fully spin-polarized.
Furthermore, the condition of connectivity for $N$-component fermions is presented, and Nagaoka's theorem is also
generalized to $SU(N)$-symmetric fermion systems on non-separable graphs.
\end{abstract}
\maketitle

{\it Introduction.} -- The origin of itinerant ferromagnetism based on Fermi surface splitting rather than the ordering of local spin moments is a difficult question in condensed matter physics.
As illustrated by Stoner's criterion, itinerant FM arises from Fermi statistics -- parallel spin alignment leads to the antisymmetrization
of electron spatial wavefunctions, which reduces the repulsive
interaction energy \cite{Stoner1938}.
However, spin polarization suffers from a kinetic energy cost,
which often dominates the gain of the exchange energy.
As a result, electrons typically remain unpolarized even in the
presence of strong interactions, developing highly correlated
wavefunctions to reduce interaction energy \cite{Lieb1962}.
Hence, non-perturbative results and exact theorems in particular are
desired for the study of itinerant FM to set up reliable benchmarks.
Known theorems of itinerant FM include Nagaoka's theorem \cite{Nagaoka1966}
and its various generalizations \cite{Tasaki1989,Tasaki1998,Li2014,
Li2015} and flat-band
ferromagnetism \cite{Tasaki1992,Mielke1992}.
Inspired by the orbital activity characterized by Hund's coupling in
most FM metals, a set of theorems of itinerant FM in orbital band
systems driven by Hund's coupling have been recently proven \cite{Li2014},
identifying phases of FM with a large range of electron fillings
and finite band width \cite{Chen2013a,Aron2014,Xu2015,Iaconis2016}.

Nagaoka's theorem, the first exact result showing itinerant FM \cite{Nagaoka1966}, proves the existence and uniqueness, up
to spin degeneracy, of the fully polarized ground state for the single-band Hubbard model.
It applies for the case with a single hole away from half-filling
in the limit of $U\to\infty$, in which the only energy is the hole's kinetic energy.
Intuitively, the hole's motion is fully coherent in the background
of a fully polarized spin configuration, while it becomes
incoherent if spins are unpolarized. Hence, the kinetic
energy is optimized with a configuration of the maximum total spin.
The proof of Nagaoka's theorem was simplified by Tasaki \cite{Tasaki1989} through use of the
Perron-Frobenius theorem, which has two key conditions---non-positivity and
connectivity.
Non-positivity means that all the off-diagonal matrix elements
of the many-body Hamiltonian are negative or zero, which is
feasible for a single hole under a suitably defined basis but
generally not for more than one hole due to fermionic statistics.
Connectivity means that the hole's motion can connect all configurations
of spins and holes.

The connectivity condition is typically difficult to verify on a general lattice.
It holds on lattices composed of loops of three or four sites
if the lattice remains connected after removing any single site
\cite{Tasaki1989,Tasaki1998}.
In this case, the hole's hopping around each loop generates arbitrary permutations of spins.
The 2D square and triangular lattices and the 3D cubic lattices satisfy
this condition, and Nagaoka's theorem applies to them.
However, for lattices consisting loops of more than 4 sites, such as
the 2D honeycomb lattice and 3D diamond lattice, it remains unclear
from previous work whether Nagaoka's theorem holds.
It is thus interesting to ask whether necessary and sufficient conditions can be determined under which connectivity is satisfied.

Graph theory has been a useful tool in organizing and solving physical problems.
A celebrated example is the diagrammatic expansion of field theory, in which graph theory is used to guide the loop expansion and the one-particle irreducible vertex expansion \cite{peskin1996}.
In the $1/N$ expansion of the large $N$-method, Feynman diagrams
are sorted based on their degree of planarity, and the leading-order contribution comes from the planar diagrams \cite{thooft2002}.
Graph theory also plays an important role in the study of polymer configuration \cite{forsman1976}, phase transitions in Ising
and Potts models \cite{essam1971}, and electric network designs \cite{estrada2013}. 

\begin{figure}
\subfloat[]{%
  \includegraphics[width=.32\linewidth]{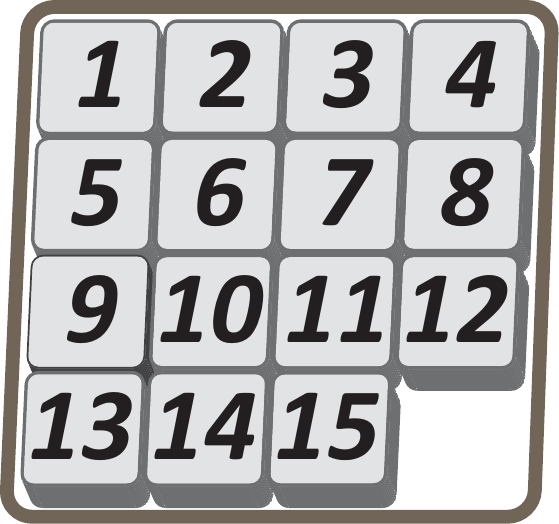}%
}
\subfloat[]{
  \includegraphics[width=.32\linewidth]{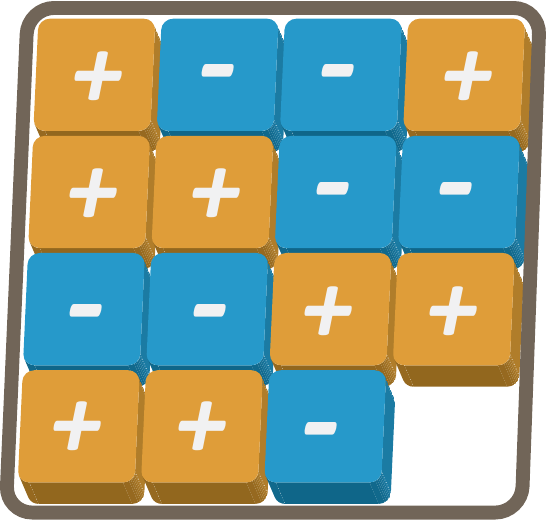}%
}
\subfloat[]{
  \includegraphics[width=.32\linewidth]{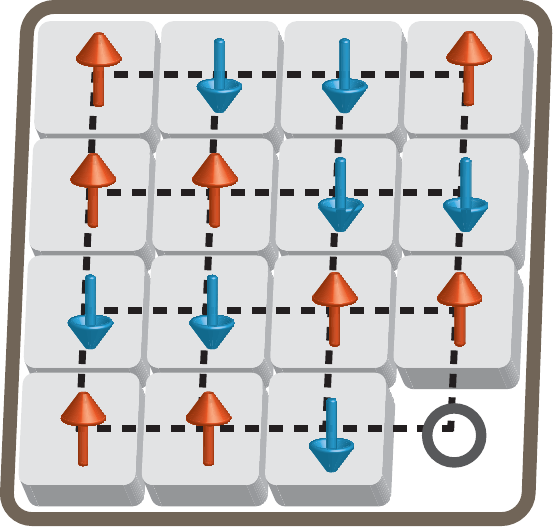}%
}
\caption{(a) The solved configuration of the original 15 puzzle. The goal of the puzzle is to return to this configuration from any scrambled starting one.
(b) For the 15-puzzle analogous to spin-$1/2$ particles,
there are only two labels.
A sample configuration on a $4\times 4$ grid is shown here
with $+$ for spin-up and $-$ for spin-down. It is mapped to a spin configuration with a single hole in a square lattice in (c). }
\label{fig:15puzzle}
\end{figure}

In this article, we find an interesting connection between the study of itinerant ferromagnetism and the celebrated 15-puzzle problem of graph theory.
In its original form, the 15-puzzle consists of a $4\times 4$ grid of tiles numbered from 1 to 15, with the 16th cell on the grid being the hole.
The hole can be transposed with neighboring tiles, and the goal is to
permute a scrambled configuration to put the tiles in order,
as shown in Fig. \ref{fig:15puzzle}.
The generalized version of the 15-puzzle problem was examined on arbitrary graphs in Ref. [\onlinecite{Wilson1974}].
By relating the connectivity condition of lattices to the 15-puzzle problem, we find 
that connectivity holds for spin-$\frac{1}{2}$ electrons {\it if and only if}
the lattice (graph) is non-separable and not a single polygon larger than a quadrilateral.
This generalizes Nagaoka's theorem to a large class of lattices including the honeycomb lattice and the diamond lattice for which
Nagaoka's theorem has not been previously proven.
We also provide criteria for the connectivity condition for
SU($N$) fermions in the fundamental representation, leading to a generalized SU($N$) Nagaoka's theorem.

In what follows, we refer to a``graph" instead of a ``lattice" since the
results require a finite number of sites and do not
depend on a regular lattice structure.
Consider a spin-$\frac{1}{2}$ Hubbard model on a general graph,
\bea
\label{eq:hamiltonian}
H = \sum_{i, j, \sigma} t_{ij} c^\dagger_{i\sigma} c_{j\sigma} + U \sum_i n_{i\uparrow} n_{i\downarrow},
\eea
where $\sigma$ is the spin index, $n_{i, \sigma} = c^\dagger_{i\sigma}c_{i\sigma}$, and $t_{ij}$ is a symmetric matrix of hopping amplitudes that encodes the graph structure.
If sites $i$ and $j$ are connected, then $t_{ij}>0$, otherwise $t_{ij}=0$.
In the limit of $U \to \infty$, states with doubly occupied sites are projected out, and every site has exactly one electron apart from
the site with a hole.
On a bipartite graph, the overall sign of $t_{ij}$ 
is not physically meaningful, since it can by changed by a gauge transformation $c_{i\sigma} \to -c_{i\sigma}$ on all sites $i$ within one of the two subgraphs.

In order to consider a general graph structure, we now summarize Tasaki's proof of Nagaoka's theorem \cite{Tasaki1989}.
Since the Hamiltonian of Eq. \eqref{eq:hamiltonian} is SU(2) symmetric,
the Hilbert space decomposes into sectors labeled by the $z$-component of total spin $S_{z,tot}$.
Without loss of generality, consider the sector where $S_{z,tot}=0$
or $1/2$ for cases with an even or odd number of spins, respectively,
since any SU(2) multiplet has a component in this sector.
The basis is defined as
\bea
|h, \{\sigma\} \rangle = (-1)^h \sideset{}{'}\prod_i c^\dagger_{i, \sigma_i}(\textbf{r}_i)|0\rangle,
\label{eq:basis}
\eea
where $c^\dagger_{i,\sigma_i}$ is ordered following an arbitrary
but fixed sequence of the vertex indices, $h$ is the index
of hole's location, and the primed product excludes the creation
operator at the hole's vertex.
In this basis, the Hamiltonian matrix satisfies
a non-positivity condition in that its elements are all $0$ or $-t_{ij}$.
Suppose that the Hamiltonian additionally satisfies a connectivity condition, which requires that
there exist a positive integer power $N$ for any two basis elements $|h, \{\sigma\}\rangle$ and $|h', \{\sigma'\}\rangle$ such that
\bea
\langle h', \{\sigma'\} | H^N | h, \{\sigma\} \rangle \neq 0.
\label{eq:conn}
\eea
This connectivity condition intuitively means that any configuration of the spins and hole in the $S_z$ sector
can be converted into any other configuration through a sequence of hole hopping.

According to the Perron-Frobenius theorem, if both non-positivity and connectivity are satisfied, Eq. \eqref{eq:hamiltonian} has a unique ground state,
\bea
\label{eq:pf}
|\psi_g \rangle = \sum_{h, \{\sigma\}} \alpha_{h,\{\sigma\}}|h,\{\sigma\}\rangle,
\eea
with a positive-definite wavefunction, meaning $\alpha_{h,\{\sigma\}}>0$
for all states in the selected $S_z$ sector.
To determine the total spin of $|\psi_g\rangle$, a trial state $|\psi_t\rangle$
is constructed by summing over all states in the $S_z$ sector with
equal weight, $|\psi_t\rangle=\sum_{h, \{\sigma\}} |h,\{\sigma\}\rangle$.
Such a state is fully symmetric under permutation of spin configurations and is thus fully spin polarized.
Since $\avg{\psi_g|\psi_t}>0$, $|\psi_g\rangle$ shares the
same quantum numbers as $|\psi_t\rangle$, meaning the ground state must also be
fully spin polarized.

In order to determine conditions under which the connectivity condition holds, it is useful to consider the generalized 15-puzzle problem, which was examined on arbitrary graphs in Ref. [\onlinecite{Wilson1974}].
Through induction on the number of loops in the graph, it is proved that, apart from two classes of exceptions, any permutation can be performed
on a non-separable, non-bipartite graph, and any even permutation can be performed on a non-separable, bipartite graph.
Here ``non-separable" means that the graph remains path-connected if
any single vertex is removed.
The first class of exceptions consists of single polygons larger than a
triangle, and the second class consists of the so-called $\theta_0$-graph
which is a single hexagon with an extra vertex in the middle that
connects two opposite hexagon vertices as shown in
Fig. 3 
in the Supplemental Material (S. M.) I.

We can now relate the connectivity condition to the generalized 15-puzzle problem.
Each electron is labeled by ``+" or ``$-$" according to its eigenvalue
$S_z=\pm \frac{1}{2}$ and electrons of the same
label are indistinguishable.
For example, Fig. \ref{fig:15puzzle} ($b$) illustrates a $4\times 4$
lattice, in which each square plaquette represents a vertex of the
corresponding graph.
The basis elements Eq. \eqref{eq:basis} correspond to an assignment of $+$, $-$, or the hole to each location.
The connectivity condition is satisfied if any configuration of labels can be converted to any other with
the same total numbers of $+$ and $-$ by a sequence of transposing the hole with neighboring labels. On a general graph, this takes the form of the generalized 15-puzzle with only two distinct tile labels.
Based on the solution to the general 15-puzzle problem \cite{Wilson1974},
we have the following theorem.


\begin{theorem}
\label{thm:1}
The Hamiltonian in Eq. \eqref{eq:hamiltonian} on a graph G satisfies the connectivity condition of Eq. \eqref{eq:conn} if and only if
G is non-separable and G is not a polygon with
$V\ge 5$ vertices.
The ground state of the model in Eq. \eqref{eq:hamiltonian}
is then fully spin-polarized and unique up to spin degeneracy
when $U \to \infty$ and there is exactly one hole.
\end{theorem}

\vspace{-2mm}
\proof
We first prove sufficiency.
The connectivity condition can be verified
if G is a single triangle or quadrilateral simply by cycling the hole
around the loop and noting that at least two spins are identical in the quadrilateral case since there are only two distinct spin labels.
Connectivity also holds on the $\theta_0$-diagram, as
shown in S. M. I.
For the remaining non-separable, non-polygonal graphs, we note that since spin only has two labels, the permutations of the spins and
hole are a subset of the possible permutations in the corresponding 15-puzzle problem
where every vertex has a distinct label.
Hence, for the non-bipartite graphs, where all permutations
can be performed in the 15-puzzle problem with all labels distinct, the connectivity condition is immediately satisfied.
For the bipartite graphs, the vertex number is larger than 4, and there are thus at least two vertices occupied by the same spin label.
Since exchanging these two 
identical spin labels 
is an identity operation, any odd permutation can be composed with an exchange of two identical spin labels to produce an even permutation with the same effect on the labels. 
Hence generating all the even permutations on the bipartite
graph is equivalent to generating all permutations when there is a repeated label. Since the 15-puzzle problem on the bipartite graphs allows for any even permutation, connectivity is satisfied.

Next we prove necessity by demonstrating that the connectivity condition is satisfied for neither polygons with vertex number $V\ge 5$ nor for separable graphs.
For the polygons with $V \ge 5$, permutations leaving the hole fixed are cyclic permutations, all generated by a single $V-1$-cycle, on the spin labels and thus cannot connect all configurations in the $S_z$ sector in general since these cyclic permutations cannot exchange neighboring spins unless every spin but one has the same label.
This can also be seen by counting the number of configurations in the $S_{z,tot}=0$ or $1/2$ sector, where the total configuration number is $N_c=V!/[m!(V-m-1)!]$ where $m=\frac{V-1}{2}$, or, $\frac{V-2}{2}$ for odd or even $V$, respectively.
Cycling the hole around the polygon can at most generate $V(V-1)$
configurations, which is less than $N_c$ for $V\ge 5$.
For the separable graphs, we only need to consider a general connected
but separable graph, which can be divided into two subgraphs $A$ and $B$ with
a single vertex $O$ connecting them. $A$ and $B$ are thus disconnected and the overall graph can be disconnected by removing $O$.
If the hole is initially at $O$, then if the hole
moves to $A$, it cannot enter $B$ without passing $O$, and vice versa.
As a result, the hole's motion cannot be used to move spins between $A$ and $B$, and
permutations can only be performed within $A$ and  $O$,
or $B$ and $O$, but not between $A$ and $B$.

\begin{figure}
\subfloat[]{%
  \includegraphics[width=.25\linewidth]{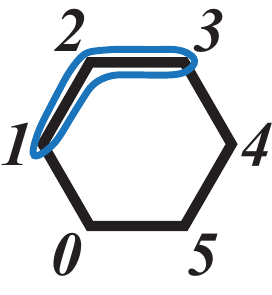}%
}\hspace{.06\linewidth}
\subfloat[]{
  \includegraphics[width=.25\linewidth]{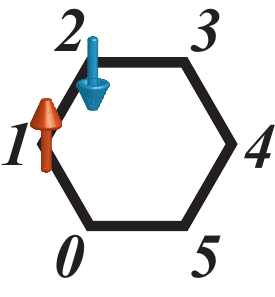}%
}\hspace{.06\linewidth}
\subfloat[]{
  \includegraphics[width=.25\linewidth]{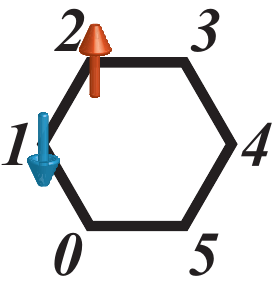}%
}
\caption{(a) A 3-cycle for any adjacent three vertices $1$, $2$,
and $3$ can be performed in a hexagon loop with concrete
steps given in S. M. II.
(b) Vertices $1$ and $2$ are occupied by spin-$\uparrow$ and $\downarrow$.
Apply $(123)$ or $(132)$ when $3$ is $\uparrow$ and $\downarrow$, respectively,
then spins on $1$ and $2$ are exchanged without affecting other vertices as shown in (c).
}
\label{fig:honeycomb}
\end{figure}


Theorem 1 ensures Nagaoka ferromagnetism for all regular
lattices, which goes beyond the previous results in literature applying for graphs composed of triangles and quadrilaterals \cite{Tasaki1989,Tasaki1998}.
For example, this demonstrates Nagaoka ferromagnetism on lattices where the minimal loops are hexagons.
To our knowledge, this has not been previously proven in spite of
the numerical evidence from  density matrix renormalization group
simulations \cite{Liu2012a}.
We thus have the following corollary. \\
{\bf Corollary}
{\it Nagaoka ferromagnetism applies to both the 2D honeycomb lattice and the 3D diamond lattice.}

We can explicitly demonstrate connectivity in the honeycomb lattice
using 3-cycles, and the same method applies to the diamond lattice as well.
Fig. \ref{fig:honeycomb} ($a$) presents that a 3-cycle for any
three adjacent vertices in a hexagon loop can be performed
for the 15-puzzle problem without affecting other vertices,
and a construction of such a 3-cycle is given in S. M. II.
For the case of spins, two opposite labels on any edge can be exchanged without affecting other vertices
as shown in Fig. \ref{fig:honeycomb} ($b$).
Without loss of generality, assume vertices 1 and 2 are
occupied with spin labels $\uparrow$ and $\downarrow$,
respectively.
If site 3 has spin-$\downarrow$, then simply applying the cycle (123)
will be exchange the spins at sites 1 and 2. Otherwise, if site 3 has spin-$\uparrow$, performing the cycle twice, or (321), will exchange the spins.
Next, consider any two vertices $1^\prime$ and $2^\prime$
with opposite spin labels.
We can choose a path connecting them.
If the hole is not on the path, it is straightforward to show
that by successively applying exchanges between adjacent
vertices along the path can exchange $1^\prime$ and $2^\prime$ without affecting other vertices.
If the hole is on the path, move it away, and after the exchange is performed, reverse the hole's motion.
Since all the permutations of spins can be generated by
exchanges, they can also be performed. In other words, the 3-cycles of adjacent vertices generate all 3-cycles on the lattice.
This establishes connectivity on the honeycomb
lattice.

Below we examine the stability of the fully polarized ferromagnetic
state in the presence of multiple holes, and the hole number is
denoted as $N_h$.
According to the Gershgorin circle theorem, the ground state
energy $E_g$ satisfies $E_g\ge \mbox{min}_l \{H_{ll} -
\sum_{m\neq l} |H_{lm}|\}$, where $l,m$ refer to the
many-body bases of Eq. \eqref{eq:basis} \cite{Tian1991,Shen1993a}.
As $U\to \infty$, the diagonal matrix elements vanish.
The configurations minimizing the contribution from the off-diagonal matrix elements are those in which no two holes are adjacent,
allowing each hole to hop to three different neighboring sites, and
thus, $E_g \ge -3 t N_h$.
To estimate an upper bound of $E_g$, we define the reference state
$|R_\uparrow\rangle$ as the fully polarized half-filled state,
where each site is occupied by one spin-$\uparrow$ electron.
A trial state $|\psi_t\rangle$ is constructed as a Fermi sea
of $N_h$ holes by removing $N_h$ electrons from the highest filled single-particle states in $|R_{\uparrow}\rangle$ as
$
|\psi_t\rangle=\prod_{i=1}^{N_h} d_{\uparrow}(\mathbf{k_i}) |R_{\uparrow}\rangle,
$
where $d_{\uparrow}(\mathbf{k_i})$ is the band eigen-operator.
The highest single-particle states correspond to the momentum $\mathbf{k_i}$ near $0$ at the top of the upper band.
Consider the limit of $N_h/M\to 0$, with $M$ the lattice site
number, in which the dispersion is parabolic as $\epsilon_k \approx 3t(1-k^2/4)$.
Since $|\psi_t\rangle$ is fully polarized, there is no double
occupancy and $\avg{H_U}=0$.
The kinetic energy is estimated as
$
E_K = -3N_h t + tM\int_0^{k_0} \frac{3k^3 dk  }{8\pi}
= - 3n_h t + t O(N_h^2/M),
$
where the Fermi wavevector $k_0$ is determined by $\frac{\pi k_0^2}{(2\pi)^2}\approx L_h/M$.
Hence, the ground state energy satisfies the bounds of
\bea
-3N_ht \le E_g \le  -3N_ht + t O(N_h^2/M).
\eea
In the thermodynamic limit, the upper and lower bounds coincide as long as $N_h$ grows more slowly than $M^\alpha$ with $
\alpha=\frac{1}{2}$,
and the fully spin-polarized trial state is a ground state.
By a similar method, it can be found that in the diamond lattice
the fully polarized state remains a ground state when
$\alpha=\frac{2}{5}$.
These values of $\alpha$ agree with the previous works for
the square and cubic lattices, respectively.

{\it Extensions} --
Recently, SU($N$) symmetric fermionic systems have attracted
considerable attention in the context of cold atom physics, where they
can be realized by alkaline earth fermions \cite{Wu2003,Wu2012}.
Consider the SU($N$) Hubbard model
\bea
H=\sum_{ij,\alpha=1}^N t_{ij} c^\dagger_{i,\alpha} c_{j,\alpha}
+\frac{U}{2}\sum_i n_i (n_i-1),
\label{eq:sun}
\eea
where $1\le \alpha \le N$ labels the fermion component, $n_i$ is the number operator
$n_i=\sum_\alpha c^\dagger_{i,\alpha} c_{i,\alpha}$, and
$t_{ij}>0$ for connected sites $i$ and $j$ with $t_{ij} = 0$ otherwise.
In the $U\to \infty$ limit 
with one hole away from $1/N$ filling, where every site but one has exactly one fermion,
Nagaoka's theorem was previously generalized to this SU($N$) system on triangular, Kagome, and
hypercubic lattices \cite{katsura2013}.
Without loss of generality, below we only consider the case
where $N$ is less or equal to the particle number, i.e., $N\le V-1$. The fermions considered here are in the fundamental representation of SU($N$), yielding $N$-component fermions.

It is natural to generalize Nagaoka's theorem to the SU($N$)
case on general graphs with the help of the 15-puzzle problem.
The non-positivity of the Hamiltonian matrix of Eq. \eqref{eq:sun}
can be established under a many-body basis constructed similarly to Eq. \eqref{eq:basis}.
For the connectivity condition, consider the non-separable graphs
other than the $\theta_0$-graph and polygons.
For non-bipartite graphs, the connectivity condition
holds even when all the occupied vertices have different fermion components, which places no further requirements on $N$.
For bipartite graphs, since only even permutations
can be performed, 
at least two vertices must be occupied by fermions in the same component to allow a 3-cycle involving two identical fermions to behave as an odd permutation. Satisfying connectivity thus requires
$V\ge N+2$ for bipartite graphs.
For polygons, connectivity only holds on triangle and quadrilateral for the SU(2) case, and it does not holds
on any polygons with $V \ge 4$ for $N\ge 3$.
For the $\theta_0$ graph, connectivity holds only for $N=2$.
Summarizing the reasoning above, we have the following
theorem.
\begin{theorem}
\label{thm:2}
Consider the SU($N$) Hamiltonian Eq. \eqref{eq:sun} on a graph
with the vertex number $V\ge N+1$ in the limit of $U\to +\infty$
with a single hole with $N>2$.
The connectivity condition is satisfied on non-separable
graphs other than the $\theta_0$-graph and polygons
with $V \ge 4$, with an additional condition that $V\ge N+2$ for
a bipartite graph.
Then the ground state is in the fully symmetric one-row
SU($N$) representation and is unique up to the SU($N$)
degeneracy.
\end{theorem}

We can also generalize the ferromagnetism to hard-core bosons
with the same Hamiltonian Eq. \eqref{eq:sun}.
The Perron-Frobenius theorem together with the 15-puzzle problem can be used to prove a fully spin-polarized
ground state.
For bosons, the hopping amplitudes need to be $t_{ij}<0$ for
links to satisfy the non-positivity of the Hamiltonian matrix
elements.
As opposed to the fermion case, extension to multiple holes is possible since bosons do not suffer from the minus sign when
switching two holes, allowing non-positivity to hold.
Connectivity continues to hold for non-separable graphs
other than single polygons larger than a triangle and the $\theta_0$-graph, since a single hole can still be
used to solve the 15-puzzle.
This yields Theorem \ref{thm:3}.
\begin{theorem}
\label{thm:3}
Consider the Hubbard model of Eq. \eqref{eq:sun} for SU($N$) hard-core bosons in the $U\to +\infty$ limit on a graph G.
The connectivity condition is satisfied for any number of bosons $N_b \leq V-1$ if and only if G is a non-separable graph
other than $\theta_0$ and polygons with $V\ge 4$ with an additional
condition that $V\ge N+2$ in the case of only a single hole
in a bipartite graph. If there are at least two holes, connectivity holds if G is $\theta_0$ as well.
Then the ground state is in the fully symmetric, one-row
representation of SU($N$), which is unique up to SU($N$) degeneracy.
\end{theorem}

{\it Conclusions.} --
The graph theorem of the generalized 15-puzzle problem has been
applied to establish the Nagaoka ferromagnetic state of the infinite-$U$
Hubbard model on general graphs with a single hole away from half-filling.
We have found that for the SU($2$) case, the Nagaoka state is
the unique ground state up to spin degeneracy for all
non-separable graphs other than single polygons with
vertex number $V\ge 5$, as established by Theorem \ref{thm:1}.
This extends Nagaoka's theorem to the 2D honeycomb lattice
and the 3D diamond lattice, whose minimal loops contain 6 vertices
and are hence beyond previous results in literature.
Furthermore, Nagaoka's theorem can also be extended to the
case of a single hole in an otherwise $1/N$-filled SU($N$)
Hubbard model.
In the SU($N$) case, the result is valid on non-separable graphs other than the $\theta_0$-graph
and single polygons with an additional condition of $V\ge N+2$
for bipartite graphs, as established by Theorem \ref{thm:2}.
Similar results can also be generalized to SU($N$) hard
core boson systems.
These results are helpful for further analytic and numeric
studies of the mechanism for itinerant ferromagnetism
and searches for novel ferromagnetic states in condensed
matter and ultra-cold atom systems.

{\it Acknowledgments.} --
E. B. and Y. L. are supported by the U.S. Department of Energy, Office of Basic Science, Division of Materials Sciences and Engineering, Grant No. DE-FG02-08ER46544.

%


\end{document}